\documentclass[final,5p,times,twocolumn]{elsarticle}

\biboptions{sort&compress}

\usepackage{graphicx}
\usepackage{amssymb}

\usepackage{amsmath}
\usepackage{bm}

\graphicspath{{.}{./EPS/}}

\journal{Physica E}

\begin{document}

\begin{frontmatter}

\title{Enhanced Seebeck effect in graphene devices by strain \\ and doping engineering}

\author[abc,cde]{M. Chung Nguyen$^{*,}$}
\author[abc,cde]{V. Hung Nguyen$^{*,}$}
\author[cde]{Huy-Viet Nguyen}
\author[abc]{J. Saint-Martin}
\author[abc]{P. Dollfus}

\address[abc]{Institut d'Electronique Fondamentale, UMR8622, CNRS, Universit$\acute{e}$ Paris Sud, 91405 Orsay, France}
\address[cde]{Center for Computational Physics, Institute of Physics, Vietnam Academy of Science and Technology, P.O. Box 429 Bo Ho, 10000 Hanoi, Vietnam}

\cortext[contact]{Corresponding authors, e-mail: mai-chung.nguyen@u-psud.fr; viet-hung.nguyen@u-psud.fr.}

\begin{abstract}
In this work, we investigate the possibility of enhancing the thermoelectric power (Seebeck coefficient) in graphene devices by strain and doping engineering. While a local strain can result in the misalignment of Dirac cones of different graphene sections in the \textit{k}-space, doping engineering leads to their displacement in energy. By combining these two effects, we demonstrate that a conduction gap as large as a few hundreds \textit{meV} can be achieved and hence the enhanced Seebeck coefficient can reach a value higher than 1.4 \textit{mV/K} in graphene doped heterojunctions with a locally strained area. Such hetero-channels appear to be very promising for enlarging the applications of graphene devices as in strain and thermal sensors.
\end{abstract}

\begin{keyword}
graphene \sep deformation \sep thermoelectric effect

\end{keyword}

\end{frontmatter}

\section{Introduction}
The thermoelectric effect can be used to directly convert a temperature difference to an electric voltage and vice versa. When a conductor is connected to a hot and a cold reservoir with a temperature difference $\Delta T$, an electrical voltage $\Delta V$ is established across the conductor according to
\begin{equation}
	\Delta V = S \Delta T
\end{equation}
where $S$ is the Seebeck coefficient characterizing the thermoelectric sensitivity of the conductor. The use of materials with high Seebeck coefficient is thus one of important factors to design efficient thermoelectric generators and coolers or thermal sensors. It is also important to maximize the power factor $\sigma S^2$ where $\sigma$ is the conductivity of the material. In electronic materials in weak scattering regime, the linear response thermoelectric coefficient is given by the Mott's formula \cite{Cutler69}
\begin{equation}
	S = \frac{1}{\sigma} \frac{k_B}{e} \int \sigma(\epsilon) \frac{\epsilon - E_F}{k_BT}\left(-\frac{\partial f}{\partial \epsilon}\right) d\epsilon
\end{equation}
where $\sigma(\epsilon)$ is the energy-dependent conductivity associated to the density $n(\epsilon)$ of electrons that fill energy states between $\epsilon$ and $\epsilon + d\epsilon$, and $f(\epsilon)$ is the Fermi-Dirac distribution function with the Fermi energy $E_F$. In conventional materials, a high Seebeck coefficient is usually found in low carrier density semiconductors while a high conductivity is found in metals. The best compromise is often to use heavily-doped semiconductors where, thanks to the finite bandgap, electrons and holes can be separated and the Seebeck coefficient is not reduced by their opposite contributions.

However, since the pioneering works of Hicks and Dresselhaus \cite{Hicks93}, nanostructuring materials into low-dimensional systems are now widely investigated to enhance the thermoelectric properties. To basically understand this size effect on the Seebeck coefficient, it is convenient to start from the simplified form of (2) derived for degenerately doped materials, i.e.
\begin{equation}
	S = -\frac{\pi^2 k_B^2T}{3e} \frac{1}{\sigma}\left.\frac{\partial \sigma(\epsilon)}{\partial \epsilon}\right|_{\epsilon = E_F}
\end{equation}
This expression suggests that any effect that can enhance the energy-dependence of the conductivity should enhance the Seebeck coefficient, e.g., by enhancing the energy-dependence of the density $n(\epsilon)$ that is directly dependent on the density of states $g(\epsilon)$. Hence, compared to bulk materials, low-dimensional systems are expected to provide higher Seebeck coefficient and power factor thanks to much higher $dg(\epsilon)/d\epsilon$. For instance, it has been confirmed experimentally first in PbTe/Pb$_{1-x}$Eu$_x$Te quantum well structures \cite{Hicks96}.
\begin{figure}[!b]
	\centering
	\includegraphics[width=3.3in]{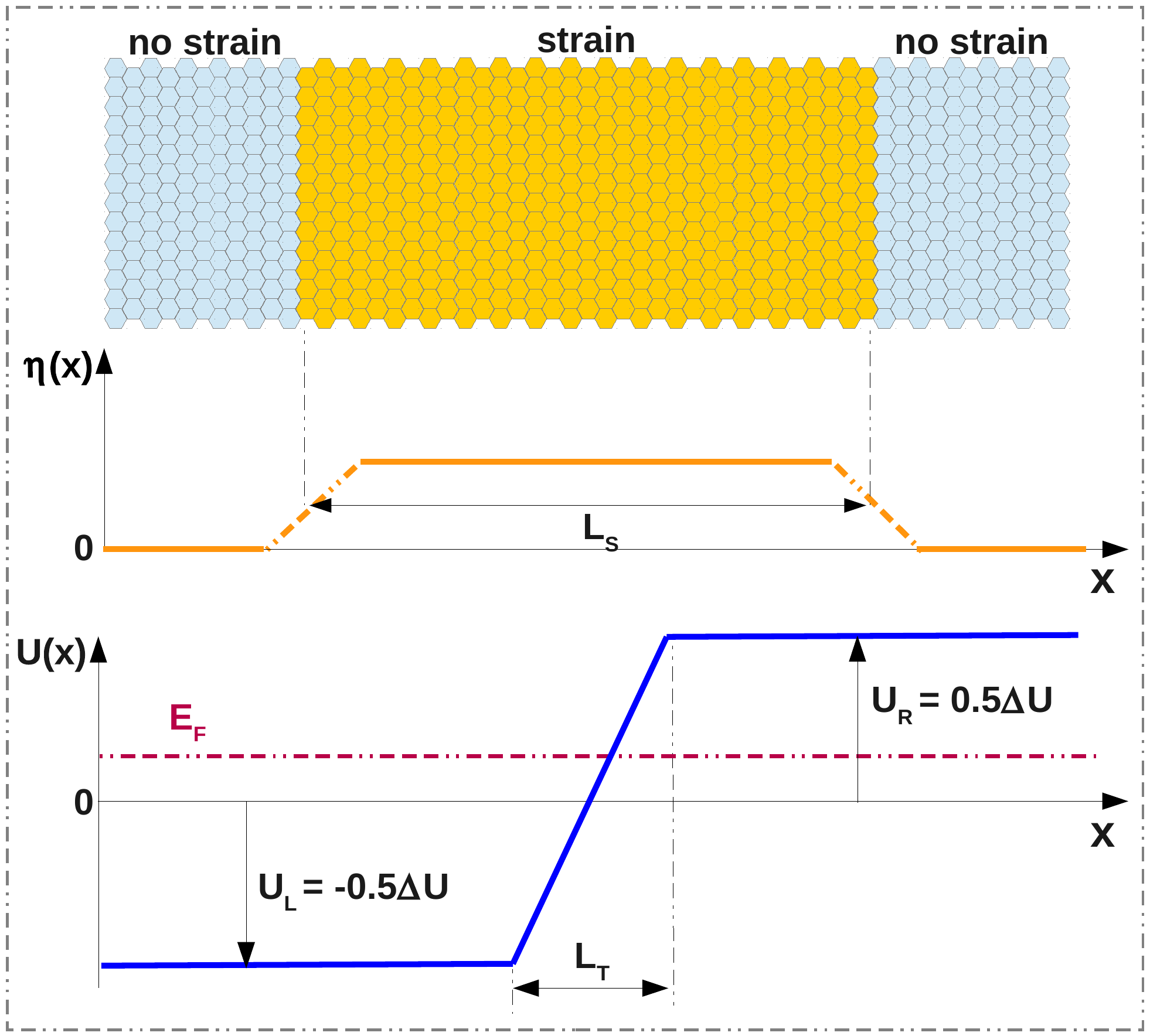}
	\caption{Schematic view of the graphene structure investigated in this work, consisting of a PN diode with a locally strained area of length $L_S$ that covers symmetrically both doped sides. The bottom shows its strain and potential profiles where the doping is characterized by the potential difference $\Delta U = U_R - U_L$ and the length $L_T$ of transition region.}
	\label{fig_sim1}
\end{figure}

The potential of graphene as thermoelectric material is quite intriguing \cite{phill15}. This single layer of carbon atoms arranged in a honeycomb lattice offers fascinating electronic properties resulting in high mobility for massless chiral particles \cite{Neto09,Bolo08}. Regarding thermoelectric properties, graphene has the advantage of a strong energy-dependence of the conductivity near the charge neutrality point \cite{Novo05}. However, it has the strong drawback to be gapless, which makes it difficult to separate the opposite contributions of electrons and holes to the Seebeck coefficient. It results in a finite but small value of $S < 100\,\mu V/K$ in pristine graphene \cite{Zuev09}.

So far, many studies have suggested different ways to open a band gap in graphene. As a direct consequence, it has been shown that the Seebeck effect can be significantly enhanced in graphene nanostructures with finite energy gap such as armchair graphene nanoribbons \cite{Zheng12}, hybrid structures combining zigzag graphene nanoribbon with zigzag boron nitride nanoribbon \cite{Yoko13}, graphene nano-hole lattices \cite{Kara11}, graphene nanoribbons consisting of alternate zigzag and armchair sections \cite{Mazz11}, vertical graphene junctions \cite{Nguyen14}, and graphene p-n junctions \cite{Shu12}. However, each mehod has its own drawbacks and still need to be confirmed by experiments.
\begin{figure}[!b]
	\centering
	\includegraphics[width=3.5in]{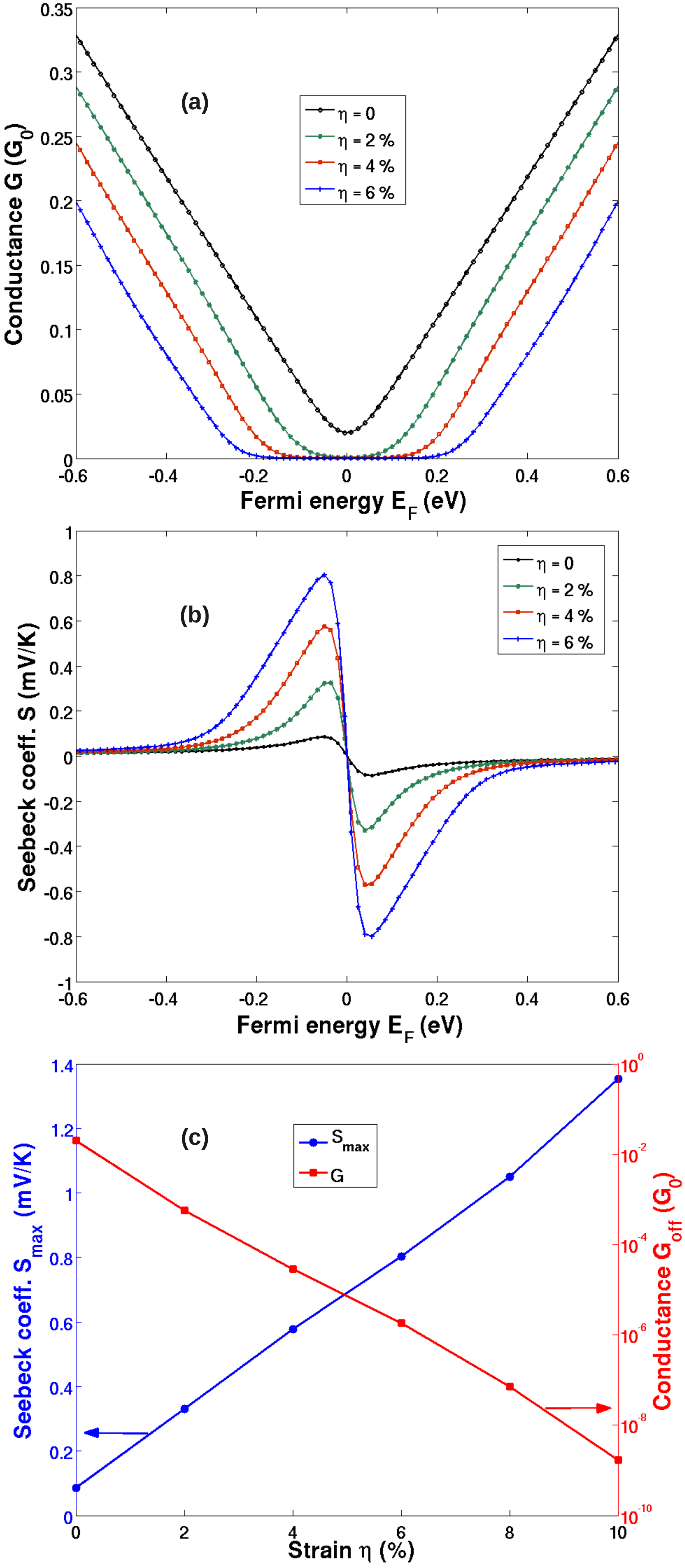} 	
	\caption{(a) conductance and (b) Seebeck coefficient as a function of Fermi energy $E_{F}$ for different strain amplitudes in the device with uniform doping. (c) maximum value of Seebeck coefficient ($S_{max}$) and conductance ($G_{off}$) in OFF state (i.e. at $E_{F} = 0$) as a function of strain amplitude.}
\end{figure}

Furthermore, graphene has been demonstrated to be conformable and able to sustain large strain \cite{shar13,Gar14,shio14}, making it a promising candidate for flexible devices. Also, strain engineering has been proposed to be an alternative approach to modulating the electronic properties of this material. In particular, it has been shown that a gap can be opened in pristine graphene for deformations beyond $20\,\%$ \cite{Pereira09}. In recent works \cite{Chung14,Hung15}, we have investigated the effects of uniaxial strain on the transport properties of 2D graphene heterochannels and found that a significant conduction gap of a few hundred meV can be achieved with a small strain of a few percent. This conduction gap is not due to a bandgap opening in the band structure but to the strain-induced shift of the Dirac cones in the Brillouin zone of different graphene sections. This result motivated us to investigate here the possible strain-induced enhancement of Seebeck coefficient in graphene nanostructures. In addition, doping engineering has been included in our investigation since it is likely to increase strongly the conduction gap, and thus the Seebeck coefficient, in graphene doped heterojunctions.

Regarding some thermoelectric applications, the figure of merit $ZT$ is another important parameter. It is defined as $ZT = \sigma S^2T/\kappa$, where $\kappa$ is the thermal conductivity. Actually, the strain engineering is not an effecient technique to modulate the phonon bandstructure \cite{gxia11} and to strongly reduce the thermal conductivity in the junctions studied here. Hence, though the Seebeck coefficient and the power factor $\sigma S^2$ are strongly improved, we believe that the combination of this design with additional nanostructuring (e.g., as in \cite{Nguyen14}) or more complex design would be required to achieve high $ZT$. For this reason, we focus here our investigation on the Seebeck coefficient that is an essential ingredient.

\section{Model and calculations}	

We investigate 2D graphene doped heterojunctions with a strain area of finite length as schematized in Fig. 1. The strain area covers symmetrically both doped sides and its length $L_S$ is assumed to be much longer than the length of the transition region $L_T$ between left and right doped sections. It has been shown that the doping profile can be generated/controlled by chemical doping or electrostatic methods, e.g., see refs. \cite{Farmer09,Williams07}. Though expected to be short for achieving high band-to-band tunneling current, the transition length is always finite in the devices with chemical doping \cite{Farmer09}. In the case of electrostatic doping \cite{Williams07}, this length is also finite but can be controlled by tuning the properties of insulator layer, i.e., its thickness and dielectric constant. Throughout this work, unless otherwise stated our calculations were performed at room temperature for $L_S = 70 \,nm$ and $L_T = 10 \,nm$.
\begin{figure}[!t]
	\centering
	\includegraphics[width=3.5in]{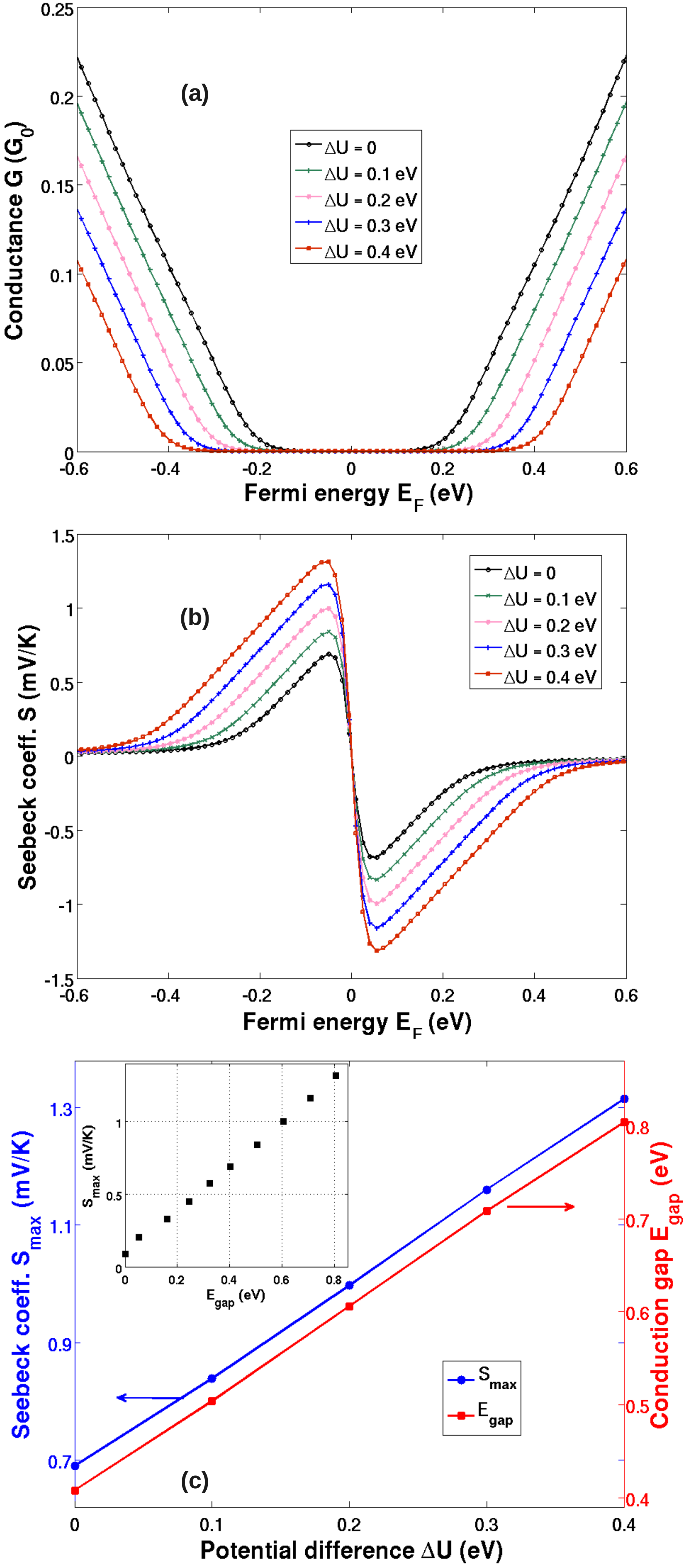} 	
	\caption{(a) conductance and (b) Seebeck coefficient as a function of Fermi energy $E_{F}$ for different $\Delta U$. (c) maximum Seebeck coefficient ($S_{max}$) and conduction gap ($E_{gap}$) as a function of $\Delta U$. Inset: $S_{max}$ as a function of $E_{gap}$. $\eta = 5\,\%$ is considered here.}
\end{figure}

\begin{figure*}[!t]
	\centering
	\includegraphics[width=6.9in]{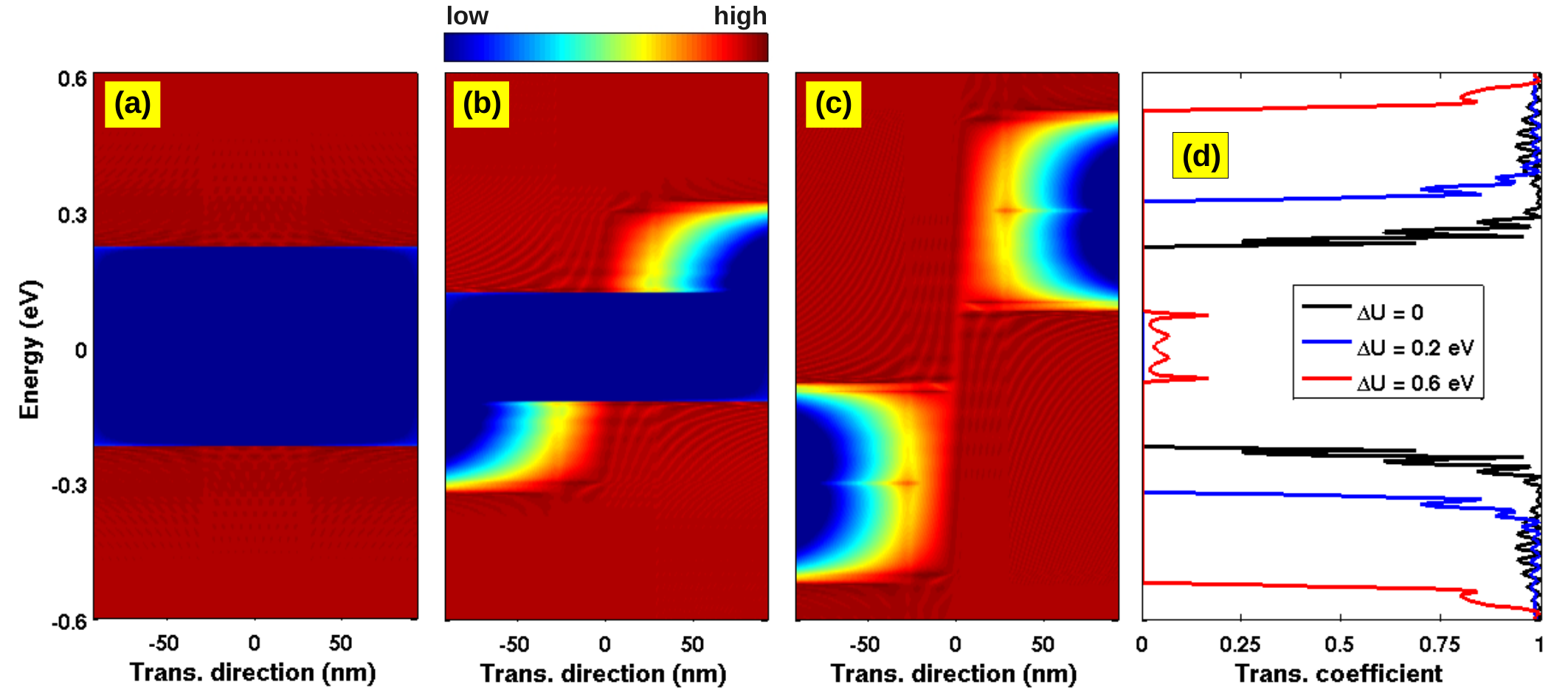}
	\caption{Local density of states in the devices of different $\Delta U$: $0$ (a), $0.2\,eV$ (b), and $0.6\, eV$ (c). The blue color regions correspond to energy-gaps, i.e., low density of states. (d) shows the transmission coefficient in these three cases. $k_y = (K^{unstrain}_y + K^{strain}_y)/2$ and $\eta = 5\,\%$ are considered here.}
\end{figure*}

A $p_z$-orbital tight-binding model was used to calculate electronic and thermoelectric properties of the device. The Hamiltonian is $H_{tb} = \Sigma_{nm} t{_{nm}} c^\dagger_{n} c_{m}$ where $t_{nm}$ is the hopping energy between nearest neighbor atoms. We consider a local uniaxial strain applied along the Oy direction. Accordingly, the strain-dependence of $C-C$ bond vectors is given by
\begin{equation}
	\left\{ \begin{array}{l}
		r_{x}\left(\eta\right) = \left(1-\eta\gamma\right) r_x\left(0\right) \\
		r_{y}\left(\eta\right) = \left(1+\eta\right) r_y\left(0\right)
	\end{array} \right.
\end{equation}
where $\eta$ is the strain amplitude and $\gamma = 0.165$ is the Poisson's ratio \cite{Blak70}. The hopping interaction between atoms is defined by $t_{nm}(\eta) = t_{0}exp{[-3.37(r_{nm}(\eta)/r_{0}-1)]}$ \cite{Pereira09}, where $t_{0} = -2.7\,eV$ and $r_{nm}(0) \equiv r_{0} = 0.142\,nm$ are the hopping energy and the \emph{C-C} distance in the unstrained case, respectively. In the specific case of this study, where the strain is applied along a zigzag axis, we have to consider two different hopping parameters $t_{1,2}$ in the armchair and zigzag directions, respectively, corresponding to two bond vectors ${\vec r}_{1,2}$ in strained graphene as in \cite{Hung14}.

This tight-binding Hamiltonian was solved by the Green's function technique \cite{Mazz12}: $\mathcal{G}\left( {\epsilon ,{k_y}} \right) = {\left[ {\epsilon  + i{0^ + } - {H_{tb}}\left( {{k_y}} \right) - {\Sigma _L}\left( {\epsilon ,{k_y}} \right) - {\Sigma _R}\left( {\epsilon ,{k_y}} \right)} \right]^{ - 1}}$, where ${H_{tb}}\left( {{k_y}} \right)$ is the Hamiltonian rewritten in the wavevector $k_y$-dependent (quasi-1D) form as in \cite{Chung14} and $\Sigma_{L(R)}$ is the self energy describing the left (right) contact-to-device coupling. The local density of states and the transmission probability needed to evaluate the transport quantities are determined as $\mathcal{D}\left( {\epsilon ,{k_y},{{\vec r}_n}} \right) =  - {\mathop{\rm Im}\nolimits} \left\{ {{\mathcal{G}_{n,n}}\left( {\epsilon ,{k_y}} \right)} \right\}/\pi$ and $\mathcal{T}_e\left( {\epsilon ,{k_y}} \right) = {\rm{Tr}}\left\{ {{\Gamma_L}\mathcal{G}{\Gamma_R}{\mathcal{G}^\dag }} \right\}$, respectively, where ${\Gamma _{L(R)}} = i\left( {{\Sigma _{L(R)}} - \Sigma _{L(R)}^\dag } \right)$ is the transfer rate at the left (right) contact. The electrical conductance and the Seebeck coefficient were calculated by
\begin{equation}
	G(E_{F}) = G_0L_{0}(E_{F},T)\\
\end{equation}
\begin{equation}
	S(E_{F}) = \frac{1}{eT}\frac{L_{1}(E_{F},T)}{L_{0}(E_{F},T)}
\end{equation}
where
\begin{equation}
	L_{n}(E_{F},T) = \frac{1}{\pi}\int d\kappa_{y} d\epsilon \mathcal{T}_{e}(\epsilon,\kappa_{y}) (\epsilon-E_{F})^{n} \left(-\frac{\partial{f}}{\partial{\epsilon}}\right).
\end{equation}
Here, $G_{0}$ = $e^{2}W/hL_{y}$ and the channel width $W = M_{cell}L_y$ with the size of unit cells $L_y$ ($\equiv r_0\sqrt{3}$) and the number of cells $M_{cell}$ along Oy direction. The integral over $\kappa_y$ ($\kappa_y \equiv k_yL_y$) is performed in the whole Brillouin zone. The distribution function $f\left(\epsilon,E_F\right) = \left(1 + \exp\left[(\epsilon-E_F)/k_BT\right]\right)^{-1}$ is the  Fermi-Dirac function with the Fermi energy $E_{F}$.

 \section{Results and discussion}

 \subsection{Strained device with uniform doping}

 First, let us examine the basic effect of strain on the transport properties of graphene. Actually, for a small strain of a few percent, graphene is still metallic \cite{Pereira09}, i.e., its gapless character do not change. However, even a small strain causes a shift of Dirac points in the \emph{k}-space \cite{Pereira09}. As a consequence, it may lead to the opening of a conduction gap in strained/unstrained graphene junctions \cite{Chung14}. This phenomenon is explained as follows. For a given $k_{y}$-mode, strained graphene has a $k_y$-dependent energy gap expressed as $E_{gap}^{strain}(k_{y})=2\left|2t_{2}\left|\cos(\frac{k_{y}L_{y}}{2})\right|-t_{1}\right|$. In the unstrained graphene, $E_{gap}^{unstrain}(k_{y})$ has a similar expression but with $t_1 = t_2 \equiv t_0$. Due to these effects of strain, the Dirac cones of unstrained/strained graphene are formed at different positions in the $k_y$-axis and hence a finite energy gap $E_{gap}^{dev}(k_{y})$ of the device transmission is always obtained for all wavevectors $k_y$. The value of $E_{gap}^{dev}(k_{y})$ corresponds to the maximum value of $E_{gap}^{unstrain}(k_{y})$ and $E_{gap}^{strain}(k_{y})$. Finally, the conduction gap is given by the minimum value of $E_{gap}^{dev}(k_{y})$ in the whole Brillouin zone, i.e,  $E_{gap}=2\left|\frac{t_{1} - t_{2}}{t_{0} + t_{2}}t_{0}\right|$ in the present case \cite{Hung14}. Besides, it has been also shown that the properties of conduction gap strongly depend on the amplitude of strain, its applied direction and lattice orientation \cite{Chung14}.

 In Figs. 2(a) and 2(b), we plot the conductance and Seebeck coefficient, respectively, as a function of Fermi energy for different strain amplitudes $\eta$ ranging from $0$ to $6\,\%$. While the minimum value of conductance is finite in pristine graphene ($\eta = 0$), a finite conduction gap is achieved when a local strain is applied to the structure. Actually, $E_{gap}$ increases almost linearly as a function of strain amplitude and, particularly, it takes the value of 0.162, 0.324 and 0.486 eV for $\eta$ $=$ $2\,\%$, $4\,\%$, and $6\,\%$, respectively. As expected from this gap enhancement, the maximum value $S_{max}$ of Seebeck coefficient increases from 0.086 mV/K for $\eta = 0$, in agreement with experimental data \cite{Zuev09}, to 0.803 mV/K for $\eta = 6\,\%$. We find as shown in Fig. 2(c) that the conductance in the OFF state, i.e., the minimum of conductance when varying $E_F$ (practically, at $E_F = 0$), decreases exponentially when the strain amplitude increases, in accordance with the linear increase of conduction gap. This effect is at the origin of the enhancement of ON/OFF current ratio in the transistors based on this type of strain heterochannels \cite{Hung14}, where the OFF and ON currents are the minimum and maximum currents, respectively, obtained when tuning the gate voltage. It is shown concomitantly in Fig. 2(c) that $S_{max}$ increases linearly as a function of strain amplitude and reaches up to \emph{1.35 mV/K} for $\eta = 10\,\%$. However, practically it may be difficult to apply locally such a large strain in this type of structure. Hence, we propose here to introduce appropriate doping engineering in this device to enhance the conduction gap with a reasonable strain amplitude.

 \subsection{Devices with both strain and doping engineering}

 \begin{figure}[!b]
 	\centering
 	\includegraphics[width=3.5in]{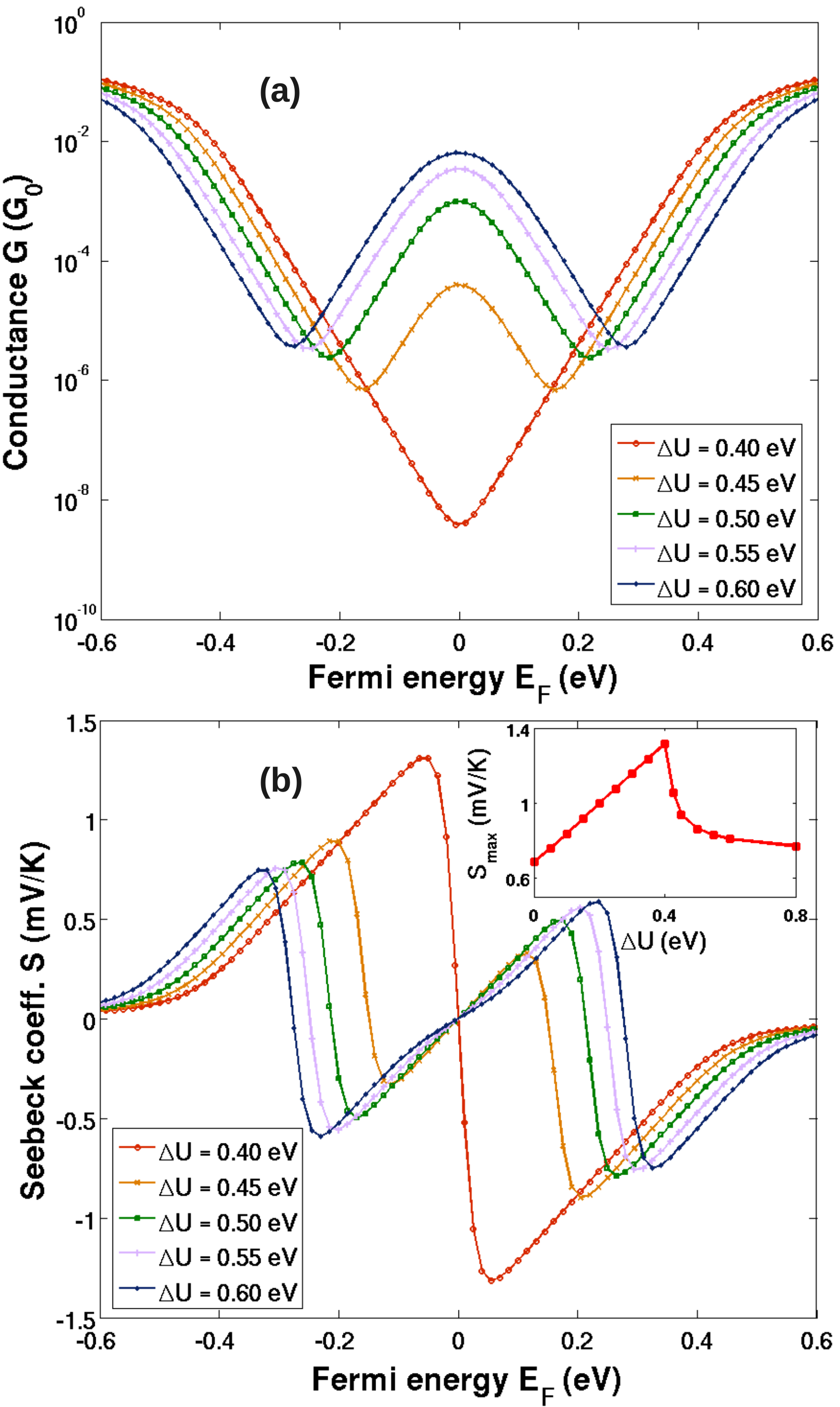}
 	\caption{(a) conductance and (b) Seebeck coefficient as a function of Fermi energy $E_{F}$ with high values of $\Delta U$. $\eta = 5\,\%$ is considered here.}
 \end{figure}
 In this subsection, we discuss the effects of doping engineering schematized in Fig. 1 on the transport properties of this strain heterostructure. The structure now consists of three sections: different doped-graphene sections in both sides and a transition region between them. Note that within the condition $L_S >> L_T$, left and right doped sections are actually formed by two different strain junctions. In Fig. 3, we display the conductance $G$ and the Seebeck coefficient as a function of Fermi energy $E_{F}$ for different doping profiles. The strain amplitude is fixed to $5\,\%$ and the doping profile is characterized by a finite potential difference $\Delta U = U_R - U_L$ (see Fig. 1). The doping engineering consists here in controlling the carrier density profile, which is characterized and determined by both the potential $\Delta U$ and the Fermi level $E_F$. The results in Fig. 3(a) show that for small $\Delta U$, the conduction gap increases with this potential difference, which can be explained as follows. When $\Delta U$ increases, $E_{gap}$ of strained/unstrained junctions in the left and right sides are shifted in opposite directions, which results in the enlargement of the transmission gap (i.e., conduction gap). Indeed, this phenomenon is clearly illustrated in the pictures of local density of states (LDOS) and transmission coefficient in Fig. 4, obtained for $k_y = (K^{unstrain}_y + K^{strain}_y)/2$ where $K^{unstrain/strain}_y$ is the wavevector at the Dirac points of unstrained/strained graphene, respectively. Note that at this $k_y$ point, $E_{gap}^{unstrain}(k_{y}) \simeq E_{gap}^{strain}(k_{y}) \simeq E_{gap}$ for $\Delta U = 0$. Actually, the conduction gap $E_{gap}(\Delta U)$ of the whole device in the range of small $\Delta U$ (i.e., $\Delta U < E_{gap}(0)$) is basically determined as $E_{gap}(\Delta U) = E_{gap}(0) + \Delta U$, as shown in Fig. 4(d) and confirmed in Fig. 3(c) where we find the linear dependence of $E_{gap}$ as a function of $\Delta U$. As a consequence, $S$ is significantly enhanced when increasing $\Delta U$ (see Fig. 3(b)) and $S_{max}$ reaches the value of 0.67, 0.84, 1.0, 1.16 and 1.32 mV/K for $\Delta U=0$, 0.1, 0.2, 0.3 and 0.4 eV, respectively. It is worth noting here that the result obtained for $\eta = 5\,\%$, $\Delta U=0.4\,eV$ is almost similar to that obtained for $\eta = 10\,\%$, $\Delta U=0$, which is about 15 times greater than $S_{max}$ in pristine graphene. Thus, it is demonstrated that the doping engineering can be an effective way to further enlarge $E_{gap}$ without the requirement of too large strain. All the features above are clearly summarized in Fig. 3(c). In addition, the inset of Fig. 3(c) confirms that whatever the technique of energy gap opening $S_{max}$ always depends linearly on $E_{gap}$, as predicted theoretically in ref. \cite{Yoko13}.

 \begin{figure}[!b]
 	\centering
 	\includegraphics[width=3.5in]{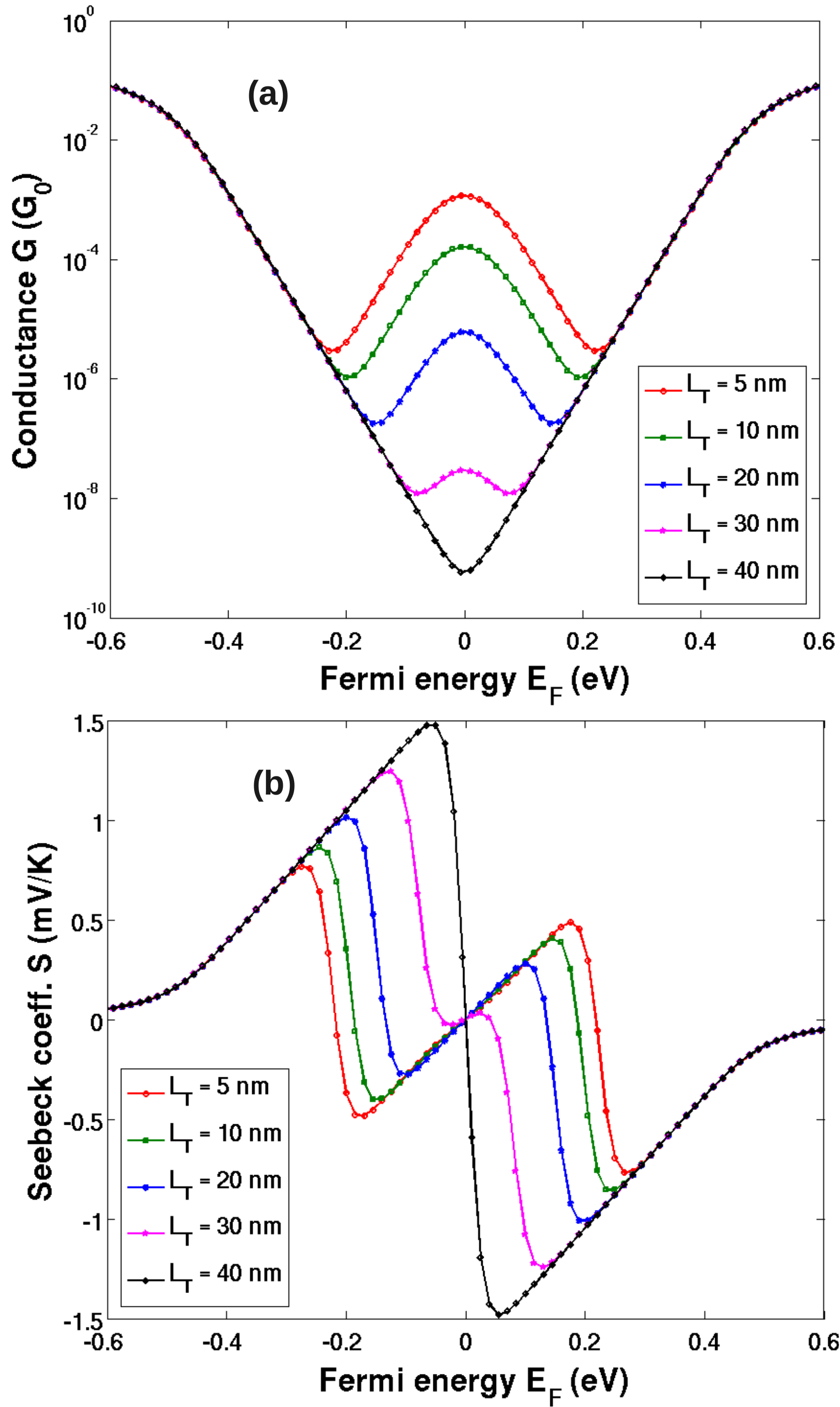}
 	\caption{(a) conductance and (b) Seebeck coefficient as a function of Fermi energy $E_{F}$ for the different lengths $L_T$. $\eta = 5\,\%$ and $\Delta U = 0.5\,eV$.}
 \end{figure}
Next, we go to analyze the effects of large $\Delta U$, i.e., values greater than $E_{gap}(0)$. In Figs. 5(a) and 5(b) we plot the conductance and Seebeck coefficient as a function of $E_F$ for large values of $\Delta U$ increasing from 0.4 eV to 0.6 eV. It is shown that the conduction gap is separated in two smaller ones that correspond to the conduction gap of each strained/unstrained junction of the structure. Between these two gaps, a region of finite conductance is recovered due to the band-to-band tunneling (BTBT), as in a standard doped tunnel diode, which is illustrated clearly in Figs. 4(c) and 4(d) where we plot the map of LDOS and transmission coefficient, respectively, for $\Delta U = 0.6\,eV$. As a consequence, at large $\Delta U$ the Seebeck coefficient exhibits two positive (negative) peaks with a maximum value $S_{max}$ that reduces and tends finally to a finite value $S_{\infty}$ when increasing $\Delta U$, as shown in the inset of Fig. 5(b). Note that the value $S_{\infty}$ is generally higher than the value of $S_{max}$ obtained for $\Delta U = 0$ but tends to this value when the transition length $L_T$ is reduced. This is explained by the detrimental contribution of band-to-band tunneling current, which is significantly reduced when increasing $L_T$ (see further comments below). It is additionally shown that $S_{max} (\Delta U)$ has a peak at $\Delta U \equiv E_{gap}(0)$, e.g., $\Delta U \simeq 0.4\,eV$ for $\eta = 5\,\%$ here.
 
 Finally, we would like to clarify the roles of device parameters $L_S$ and $L_T$ on the obtained results. Note that in this work, we consider only the case $L_S >> L_T$. Within this condition, each doped section contains by a strain junction and hence has a finite conduction gap. In principle, this conduction gap is strongly dependent on the length of the strained graphene part in these two doped sections, i.e., the transmission probability in the gap increases exponentially when reducing the length of strained graphene area. Hence, to ensure that the transmission is fully suppressed in the gap, the length $L_S$ should be much larger than $L_T$. Moreover, the required length $L_S$ is basically dependent on the value of energy gap, i.e., the larger $L_S$ is required for the smaller $E_{gap}$ (i.e., smaller strain). In particular, in the case of $\eta = 5\, \%$, $L_S > L_T + 20 \, nm$ should be used. Additionally, it has been shown that the length $L_T$ of the transition region between $n$- and $p$-doped sections plays an important role on the BTBT current \cite{hung11}, i.e., this current is exponentially reduced when increasing $L_T$ as seen in Fig. 6(a). More interestingly, based on this reduction of BTBT current, the Seebeck coefficient in the case of $\Delta U > E_{gap}(0)$ is significantly enhanced when increasing $L_T$, i.e., $S_{max}$ reaches 1.48 mV/K for $L_T = 40 \,nm$ as shown in Fig. 6(b) while it is only about 0.77 mV/K for $L_T = 5 \,nm$. We notice that in the case of small $\Delta U$, the BTBT current is negligible as shown for $\Delta U = 0$ and 0.2 $eV$ in Fig. 4 and hence $S$ is very weakly dependent on $L_T$.

 \section{Conclusion}
 In this work, we have proposed to make appropriate use of strain and doping engineering to generate and enlarge a conduction gap in graphene heterochannels and to benefit from this feature to enhance the Seebeck effect. The maximum value $S_{max}$ of the Seebeck coefficient was shown to increase linearly with the conduction gap. Remarkably, with a small strain of $5\,\%$ and an appropriate doping profile, the Seebeck coefficient can reach a value higher than 1.4 mV/K, i.e., 17 times higher than the value in gapless pristine graphene. Besides its use in strain sensors, this design strategy is thus promising to achieve good performance in graphene devices based on the Seebeck effect, as thermal sensors.

 \textbf{Acknowledgments.} This research in Hanoi is funded by Vietnam's National Foundation for Science and Technology Development (NAFOSTED) under grant number 103.01-2014.24.

\end{document}